\documentclass[12pt]{article}
\usepackage{amsthm,amsfonts, amsmath}
\theoremstyle{plain}
\newtheorem{thm}{THEOREM}[section]

\theoremstyle{definition}
\newtheorem{defi}[thm]{DEFINITION}
\theoremstyle{remark}

\newcommand{\upchi}{\raise1pt\hbox{$\chi$}}
\newcommand{\R}{{\mathord{\mathbb R}}}
\newcommand{\C}{{\mathord{\mathbb C}}}
\newcommand{\Z}{{\mathord{\mathbb Z}}}

\newcommand{\id}{{\mathord{\;\rm d}}}

\newcommand{\mfr}[2]{{\textstyle\frac{#1}{#2}}}
\newcommand{\const}{C}

\newcommand{\an}{{\mathord{a}}^{\phantom{*}}}

\newcommand{\cA}{{\mathord{\mathcal A}}}

\newcommand{\cF}{{\mathord{\mathcal F}}} 
 
\newcommand{\cK}{{\mathord{\mathcal K}}}
\newcommand{\cN}{{\mathord{\mathcal N}}}
\newcommand{\cH}{{\mathord{\mathcal H}}}
\newcommand{\cO}{{\mathord{\mathcal O}}}
\newcommand{\cU}{{\mathord{\mathcal U}}}
\newcommand{\cV}{{\mathord{\mathcal V}}}
\newcommand{\Tr}{{\mathord{\hbox{Tr}}}}

\begin{document}
\title{Upper Bounds to the Ground State Energies of the One- and
  Two-Component Charged Bose Gases} \author{Jan Philip
  Solovej\thanks{Work partially supported by NSF grant DMS-0111298, by
    EU grant HPRN-CT-2002-00277, by MaPhySto -- A Network in
    Mathematical Physics and Stochastics, funded by The Danish
    National Research Foundation, and by grants from the Danish
    research council.
    }\\
  \normalsize Institute for
    Mathematical Sciences\thanks{Most of this work was done
while the author was visiting the School of Mathematics, Institute for Advanced
Study, Princeton. \newline \copyright 2004 by the
    author. This article may be
    reproduced in its entirety for non-commercial purposes.}
   \\ \normalsize University of
   Copenhagen\\\normalsize 
 Universitetsparken 5\\
    \normalsize DK-2100 Copenhagen\\ \normalsize DENMARK. \\
  {\it e-mail\/}: solovej@math.ku.dk} \bigskip
\date{}
\maketitle 

\begin{abstract}We prove upper bounds  on the ground  state energies of the
  one- and  two-component charged Bose gases.  The upper bound for the
  one-component gas agrees   with the high density asymptotic  formula
  proposed by L.~Foldy in 1961. The upper  bound for the two-component
  gas agrees in the  large particle number  limit with  the asymptotic
  formula conjectured  by F.~Dyson in  1967. Matching asymptotic lower
  bounds  for these  systems were  proved in  references \cite{LS} and
  \cite{LS2}.  The formulas of Foldy and Dyson which are based on
  Bogolubov's pairing theory have thus been validated.
\end{abstract}

\section{Introduction and main results}

In 1961 L.~Foldy \cite{Foldy} used Bogolubov's 1947 pairing theory
\cite{BO} for Bose systems to give a heuristic calculation of the
ground state energy of a one-component charged Bose gas in the high
density limit. The one-component Bose gas is a system of Bose
particles all of the same charge moving in the presence of a fixed
uniform background of the opposite charge.

In 1967 F.~Dyson \cite{Dyson} considered the two-component Bose gas
with two species of bosons with opposite charges. Motivated
by Foldy's calculation Dyson was able to prove a rigorous upper bound
on the ground state energy. A famous consequence of Dyson's upper
bound is that charged bosonic matter is not stable, the ground state
energy is super-linear in the number of particles.
Dyson, moreover, conjectured an exact asymptotic form of the ground state energy 
in the limit of a large number of particles. 

In \cite{LS} it was proved that Foldy's calculation is indeed correct
as a leading asymptotic lower bound for the ground state energy of the one-component 
charged Bose gas in the high density limit.

In \cite{LS2} it was similarly proved that Dyson's conjectured
expression is correct as an asymptotic lower bound for the ground
state energy of the two-component charged Bose gas in the limit of a
large number of particles.

The aim of the present paper is to prove the corresponding upper bounds thus 
validating both Foldy's one-component and Dyson's two-component formulas.

It should be mentioned that Foldy's calculation may be viewed as a
trial state calculation and may thus be turned into a rigorous upper
bound. Foldy, however, uses periodic boundary conditions, and a
periodic version of the Coulomb potential.  It is not known whether
this formulation has the same thermodynamic limit as the formulation
given below.

The one-component Bose gas is a system of $N$ particles all of the {\it
  same} charge $+1$, say, constrained to a box $\Lambda=[0,L]^3\subset
\R^3$, in which there is a uniform background charge of density
$\rho$. 

The Hamiltonian for the one-component charged Bose gas is thus
\begin{equation}\label{H1N}
  H_N^{(1)}=  \sum_{i=1}^{N} (- \mfr{1}{2}\Delta_i 
  -V(x_i))
  +\sum_{1 \leq i < j \leq N} |x_i - x_j|^{-1}  
  +C
\end{equation}
where
$$
V(x)=\rho\int_\Lambda|x_i-y|^{-1}\id y,\qquad
C=\frac{\rho^2}{2}\iint_{\Lambda\times\Lambda}|x-y|^{-1}\id x\id y.
$$
We use {\it Dirichlet} boundary conditions. 

It is known from the work of Lieb and Narnhofer \cite{LN} that
the ground state energy $E^{(1)}(N)$ of $H_N^{(1)}$ has a
thermodynamic limit if we restrict to a neutral system
$$
e(\rho)=\lim_{N\to\infty\atop L^3=N/\rho}\frac{E^{(1)}(N)}{L^3}.                   
$$
It is however also shown in \cite{LN} that one will get the same
thermodynamic energy by minimizing over all particle numbers, i.e., 
$$
e(\rho)=\lim_{L\to\infty}\inf_{N}\frac{E^{(1)}(N)}{L^3}.  
$$

\begin{thm}[Foldy's formula]\label{thm:foldy} 
The ground state energy $e(\rho)$ of the one-component charged Bose gas 
  satisfies the asymptotics
\begin{equation}
  \lim_{\rho\to\infty}\rho^{-5/4}e(\rho)=-I_0,
\end{equation}
where
\begin{equation}\label{eq:I0def}
  I_0=(2/\pi)^{3/4}\int_0^\infty1+x^4-x^2\left(x^4+2\right)^{1/2}\id x=
  \frac{4^{5/4}\Gamma(3/4)}{5\pi^{1/4}\Gamma(5/4)} .
\end{equation}
\end{thm}

The two component Bose gas is described by the Hamiltonian
$$
H_N^{(2)}=\sum_{i=1}^N-\frac{1}{2}\Delta_i+\sum_{1\leq i<j\leq
  N}\frac{e_ie_j}{|x_i-x_j|}.
$$
acting on the Hilbert space $L^2(\R^3\times\{1,-1\})$, where the variable 
$(x_i,e_i)\in\R^3\times\{1,-1\}$ gives the position and charge of particle $i$.   

The word {\it two component} refers to the fact that the charge of
each particle can be either positive or negative.  Thus the gas has a
positive and a negative component.  One would not normally consider
the charges as variables, but rather fix them to have given values.
If we did that, the Hamiltonian would not be fully symmetric in all $N$
variables, but only in the variables for the positively charged
particles and negatively charged particles separately. Clearly, the
charge variables commute with the Hamiltonian and the bottom of the
spectrum (the ground state energy) $E^{(2)}(N)$ of $H^{(2)}_N$ will therefore be
achieved for a fixed combination of charges (rather than a
superposition).

\begin{thm}[Dyson's formula]\label{thm:dyson} The ground state energy 
$E^{(2)}(N)$ of the two-component charged Bose gas satisfies the
asymptotics
$$
\lim_{N\to\infty}N^{-7/5}E^{(2)}(N)= -A
$$ 
where $A$ is the positive constant determined by the variational principle
\begin{equation}\label{eq:variational}
  -A=\inf\biggl\{\mfr{1}{2}\int_{\R^3}|\nabla\Phi|^2-I_0\int_{\R^3} \Phi^{5/2}\ 
  \biggr|\ 0\leq \Phi,\ \int_{\R^3}\Phi^2=1\biggr\}
\end{equation}
with $I_0$ again given by (\ref{eq:I0def}).
\end{thm}

In \cite{Dyson} Dyson proves that $E^{(2)}(N)\leq -\const N^{7/5}$,
but with a constant different from $A$. He conjectures that the
correct value is given as above. That the exponent $7/5$ is, indeed,
correct was first proved in 1988 by Conlon, Lieb, and Yau in
\cite{CLY}, where they show a lower bound $-\const N^{7/5}$, but
still not with the correct constant. They also proved that $5/4$ is the correct exponent in
Foldy's formula. The asymptotic lower bounds in
Theorems~\ref{thm:foldy} and \ref{thm:dyson} were proved in \cite{LS}
and \cite{LS2} respectively. The main results of the following paper
are the asymptotic upper bounds.

In Sect.~\ref{sec:trialstate} we give a general construction of
bosonic trial states on the bosonic Fock space over a general Hilbert
space. The trial states will be build from coherent states and squeezed
states. The trial states are essentially the ones dictated by Bogolubov 
theory. These trial states are the bosonic equivalent of the fermionic
states in Hartree-Fock theory or rather to their extension including
the Bardeen-Cooper-Schrieffer states (see \cite{BLS}). 

In the same way as fermionic systems may be approximated by the
semi-classical Thomas-Fermi theory we will also use a semi-classical 
type approximation to the Bogolubov trial states. 

In Sect.~\ref{sec:two} we use the general trial state method to give
an upper bound on the ground state energy for the two-component gas,
but in a grand canonical setting where we do not fix the total number
of particles.

In Sect~\ref{sec:fixedN} we show how to get an upper bound for fixed
particle number and thus prove Theorem~\ref{thm:dyson}.

In Sect.~\ref{sec:one} we use the general trial state method to give
an upper bound on the ground state energy for the one-component gas
and prove Theorem~\ref{thm:foldy}.

A key ingredient in the proofs is a semiclassical construction where
we represent operators as phase-space integrals with coherent states
symbols and use the Berezin-Lieb inequalities.  We need an operator
version of the inequality.  This is discussed in
Appendix~\ref{sec:appendix}.

{\bf Acknowledgment:} I would like to thank Elliott Lieb, Kumar Raman,
and Robert Seiringer for valuable discussions.

\section{The abstract trial state construction}\label{sec:trialstate}

Our goal in this section is to construct trial states on the bosonic Fock space
$\cF=\cF(\cH_1)=\bigoplus_{N=0}^\infty \cH_N$, over some Hilbert Space $\cH_1$,
i.e., $\cH_N=\bigotimes_{\rm Sym}^N\cH_1$ and $\cH_0=\C$.  We will be
using the language of bosonic creation and annihilation operators as a
convenient tool for the book keeping.  We denote by $|0\rangle$ the
vacuum vector in $\cF$. If $T$ is an operator on $\cH_1$ and $W$ is an
operator $\cH_1\otimes\cH_1$, which is symmetric under interchange of
the tensor factors, we may lift (also referred to as second quantize)
these operators to $\cF$ as 
$$ \bigoplus_{N=1}^\infty \sum_{i=1}^N T_i\quad\hbox{and}\quad
\bigoplus_{N=2}^\infty \sum_{1\leq i<j\leq N} W_{ij}.
$$ Here $T_i$ refers to the operator $T$ acting on the $i$th factor in
the tensor product and $W_{ij}$ refers to $W$ acting on the $i$th and
$j$th factors. If $u_\alpha$, $\alpha=1,\ldots$ is an orthonormal
basis for $\cH_1$ we can express these operators using creation and
annihilation operators as
\begin{equation}\label{eq:2nd1}\bigoplus_{N=1}^\infty \sum_{i=1}^N
  T_i=\sum_{\alpha,\beta}(u_\alpha,Tu_\beta)a(u_\alpha)^*a(u_\beta)
\end{equation}
and
\begin{equation}\label{eq:2nd2}
  \bigoplus_{N=0}^\infty \sum_{1\leq i<j\leq N}
  W_{ij}=\frac{1}{2}\sum_{\alpha\beta\mu\nu}(u_\alpha\otimes
  u_\beta,W u_\mu\otimes u_\nu) a(u_\alpha)^*a(u_\beta)^*a(u_\nu)a(u_\mu).
\end{equation}
Of special interest is the number operator (the second quantization of
the identity)
$$
\cN=\bigoplus_{N=0}^\infty N.
$$
 If $\phi\in \cH_1$ is a not necessarily normalized vector we define
 the corresponding {\it coherent} state as the normalized vector in
 Fock space
\begin{eqnarray}
  |\phi\rangle_{\rm C}&=&\exp(-\|\phi\|^2/2+a(\phi)^*)|0\rangle\nonumber\\&=&
  \sum_{n=0}^\infty
  e^{-\|\phi\|^2/2}\frac{(a(\phi)^*)^n}{n!}|0\rangle,\label{eq:coherent}
\end{eqnarray}
and for a normalized $\psi\in\cH_1$ we define the {\it squeezed} state depending on $\lambda\in\C$ with
$|\lambda|< 1$
\begin{eqnarray}
  |\lambda;\psi\rangle_{\rm S}&=&
  (1-|\lambda|^2)^{1/4}\exp(-(\lambda/2)
  a(\psi)^*a(\psi)^*)|0\rangle\nonumber\\
  &=&
  (1-|\lambda|^2)^{1/4}\sum_{n=0}^\infty\frac{(-\lambda/2)^n}{n!}(a(\psi)^*)^{2n}
  |0\rangle.\label{eq:squeezed}
\end{eqnarray}
It is straightforward to check that these states are normalized. Up to
an overall phase $|\phi\rangle_{\rm C}$ and $|\lambda;\psi\rangle_{\rm
  S}$ are characterized by
\begin{equation}\label{eq:coherentsqueezedcharac}
  (a(\phi)-\|\phi\|^2)|\phi\rangle_{\rm C}=0\quad
  \hbox{and}\quad (a(\psi)+\lambda a(\psi)^*)|\lambda;\psi\rangle_{\rm S}=0.
\end{equation}
We immediately see that 
\begin{equation}\label{eq:coherentexpectation}
{}_{\rm C}\langle \phi|(a(\phi)^*)^ma(\phi)^k|\phi\rangle_{\rm C}=\|\phi\|^{2(m+k)}.
\end{equation}
For the squeezed state we get
\begin{eqnarray}
  \lefteqn{{}_{\rm S}\langle\lambda;\psi|(a(\psi)^*)^ja(\psi)^{j+2k}
    |\lambda;\psi\rangle_{\rm S}}&&\nonumber\\
  &=&(1-|\lambda|^2)^{1/2}\sum_{n=0}^\infty\frac{(2n+2k)!}{(n+k)!^2}
  (2n-j+1)(2n-j+2)\cdots(2n)\nonumber\\
  &&\times(n+k)(n+k-1)\cdots(n+1)(|\lambda|/2)^{2n}(-\lambda/2)^k\nonumber\\
  &=&(1-|\lambda|^2)^{1/2}|\lambda|^j(-\lambda)^k
  \frac{d^j}{d|\lambda|^{j}}\left(|\lambda|^{-1}\frac{d}{d|\lambda|}\right)^k
  (1-|\lambda|^2)^{-1/2}.\label{eq:squeezedexpectation}
\end{eqnarray}
Moreover, the expectation in the state $|\lambda;\psi\rangle_{\rm S}$
of a product of an odd number of the operators  $a(\psi)^*$ or
$a(\psi)$ vanishes.

For the expectation of the particle number we find
$$
{}_{\rm C}\langle \phi|a(\phi/\|\phi\|)^*a(\phi/\|\phi\|)|\phi\rangle_{\rm C}=\|\phi\|^2\quad\hbox{and}\quad{}_{\rm
  S}\langle\lambda;\psi|a(\psi)^*a(\psi)|\lambda;\psi\rangle_{\rm
  S}=\frac{|\lambda|^2}{1-|\lambda|^2}.
$$
We point out that the variation in the particle number is very
different in the coherent state and in the squeezed state
\begin{eqnarray}
{}_{\rm C}\langle
\phi|(a(\phi/\|\phi\|)^*a(\phi/\|\phi\|))^2|\phi\rangle_{\rm C}-
{}_{\rm C}\langle \phi|a(\phi/\|\phi\|)^*a(\phi/\|\phi\|)|\phi\rangle_{\rm C}^2&=&
\|\phi\|^2\label{eq:coherentstd}\\
{}_{\rm
  S}\langle\lambda;\psi|(a(\psi)^*a(\psi))^2|\lambda;\psi\rangle_{\rm
  S}-
{}_{\rm
  S}\langle\lambda;\psi|a(\psi)^*a(\psi)|\lambda;\psi\rangle_{\rm
  S}^2&=&\frac{2|\lambda|^2}{(1-|\lambda|^2)^2}.\label{eq:squeezedstd}
\end{eqnarray}
Thus in the coherent state the standard deviation of the particle number is
the square root of the expectation itself,
whereas for the squeezed state the standard deviation of the particle
number is, in fact, greater than the expectation itself. 
For this reason the squeezed states are not appropriate for
describing Bose condensates with a macroscopic and
sharply defined occupation number in a specific one-particle state. To
describe condensates we will use coherent states. 

We will here define a variational principle corresponding to the Bogolubov
theory of Bose gases. We shall do this by characterizing the set of
variational trial states.  

The Bogolubov variational theory  is very similar to the 
Hartree-Fock theory for Fermi gases. 
More precisely, it is similar to the generalized Hartree-Fock theory
which includes the Bardeen-Cooper-Schrieffer (BCS) trial states.
In generalized Hartree-Fock theory (see \cite{BLS}) the class of trial
states is defined to be the quasi-free states on a fermionic Fock
space. For the ground state (zero temperature) theory we may restrict
to {\it pure} quasi-free states. 

To describe the variational states of Bogolubov theory we 
we will again start from (normalized) quasi-free pure states. 
Such a state may be characterized as follows.
If $\Psi\in\cF(\cH_1)$ is a normalized 
quasi-free pure states there exists an orthonormal family 
$\psi_1,\ldots$ of $\cH_1$ and a sequence of numbers
$0<\lambda_1,\ldots<1$ with
$\sum_{\alpha=1}^\infty\lambda_\alpha^2<\infty$
such that 
\begin{equation}\label{eq:quasi-free}
\Psi=\prod_{\alpha=1}\left[(1-\lambda_\alpha^2)^{\frac{1}{4}}
\exp\left(-\frac{\lambda_\alpha}{2}
      a(\psi_\alpha)^*a(\psi_\alpha)^*\right)\right]|0\rangle
\end{equation}
A straightforward but lengthy calculation from
(\ref{eq:squeezedexpectation}) shows that the quasi-free state satisfies
\begin{eqnarray}\label{eq:wick}
  (\Psi,a^\sharp_1a^\sharp_{2}a^\sharp_3a^\sharp_4\Psi)
  &=&(\Psi,a^\sharp_1a^\sharp_{2}\Psi)(\Psi,a^\sharp_3a^\sharp_4\Psi)+
  (\Psi,a^\sharp_1a^\sharp_{4}\Psi)(\Psi,a^\sharp_2a^\sharp_3\Psi)\nonumber\\&&+
  (\Psi,a^\sharp_1a^\sharp_{3}\Psi)(\Psi,a^\sharp_2a^\sharp_4\Psi)
\end{eqnarray}
and from the definition of the state we have for all
integers $m\geq1$
\begin{equation}\label{eq:oddexpec}
  (\Psi,a^\sharp_1\cdots a^\sharp_{2m-1}\Psi)=0.
\end{equation}
In (\ref{eq:wick}) and (\ref{eq:oddexpec}), $a_j^\sharp$, $j=1,2\ldots$ refer to any 
creation or annihilation operators. The relation (\ref{eq:wick}) is 
the case $m=2$ of the more general rule
\begin{eqnarray}
  (\Psi,a^\sharp_1\cdots a^\sharp_{2m}\Psi)=\sum_{\pi\in P_{2m}}(\Psi,a_{\pi(1)}^\sharp
  a_{\pi(2)}^\sharp\Psi)\cdots (\Psi,a_{\pi(2m-1)}^\sharp a_{\pi(2m)}^\sharp\Psi),
\end{eqnarray}
where $P_{2m}$ is the set of 
pairing permutations
\begin{eqnarray}
  P_{2m}=\{\pi\in S_{2m}\ |\ \pi(2j-1)<\pi(2j+1),\ j=1,\ldots,m-1\nonumber\\
  \pi(2j-1)< \pi(2j),\ j=1,\ldots,m
  \}.
\end{eqnarray}
We shall here use this only in the case (\ref{eq:wick}) when $m=2$.
 
The {\it one-particle density matrix} of the quasi-free state $\Psi$
is the operator  
$\gamma_1$ defined on the one-body space $\cH_1$ by 
$(g,\gamma_1 f)_{\cH_1}=(\Psi,a(f)^*a(g)\Psi)_\cF$ where
$f,g\in\cH_1$. {F}rom (\ref{eq:squeezedexpectation})
\begin{equation}\label{eq:gammalambda}
\gamma_1=\sum_{\alpha=1}^\infty
\frac{\lambda_\alpha^2}{1-\lambda_\alpha^2}
|\psi_\alpha\rangle\langle\psi_\alpha|.
\end{equation}
Note, in particular, that the 
one-particle density matrix is a positive semi-definite trace
class operator with 
$$\Tr\gamma_1=(\Psi,\cN\Psi)
=\sum_{\alpha=1}^\infty\frac{\lambda_\alpha^2}{1-\lambda_\alpha^2}<\infty.$$
Connected to the quasi-free pure state $\Psi$ we also have 
the symmetric bilinear form $\xi_1$ on $\cH_1$ given by 
$
\xi_1(f,g)=(\Psi,a(f)^*a(g)^*\Psi)_\cF.
$
We find,  again from (\ref{eq:squeezedexpectation}), that 
\begin{equation}\label{eq:alphalambda}
\xi_1(f,g)=\sum_{\alpha=1}^\infty
\frac{-\lambda_\alpha}{1-\lambda_\alpha^2} (\psi_\alpha,f)(\psi_\alpha,g).
\end{equation}

We may identify $\xi_1$ with a linear map $\xi_1:\cH_1\to\cH_1^*$,
from the one-body space $\cH_1$ to its dual space $\cH_1^*$. We then
have the relations
\begin{equation}\label{eq:gammaalpharelation}
\xi_1^*\xi_1=\gamma_1(\gamma_1+1)\qquad
\xi_1\gamma_1=\gamma_1\xi_1,
\end{equation}
where we have also identified $\gamma_1$ in the natural way 
with a map from $\cH_1^*$ to itself.
If we introduce the operator
$\Gamma:\cH_1\oplus\cH_1^*\to\cH_1\oplus\cH_1^*$ defined using 
matrix notation as
$$
\Gamma=
\begin{pmatrix}\gamma_1 &\xi_1\\\xi_1^*&1+\gamma_1
\end{pmatrix}
$$
we may rewrite the condition (\ref{eq:gammaalpharelation}) as
$$
\Gamma\begin{pmatrix}-1&0\\0&1
\end{pmatrix}\Gamma=\Gamma.
$$
We may refer to an operator satisfying this condition as a {\it
  symplectic projection}. In the fermionic case the corresponding
operator is simply a projection.  Note that the operator $\Gamma$ may
also be described by
$$
\left(|f_1\rangle\oplus\langle g_1|,\Gamma
  |f_2\rangle\oplus\langle g_2|\right)_{\cH_1\oplus\cH_1^*}
=\left(\Psi,(a(f_2)^*+a(g_2))(a(f_1)+a(g_1)^*)\Psi\right)_{\cF(\cH_1)},
$$
where we have used the Dirac bra and ket notation to denote elements
of $\cH_1$ and $\cH_1^*$ respectively. 

Given a positive definite trace class operator $\gamma_1$ and a
symmetric bilinear form $\xi_1$ satisfying
(\ref{eq:gammaalpharelation}) we may
find a unique quasi-free pure state $\Psi$ such that $\gamma_1$ is the
corresponding one-particle density matrix and
$\xi_1$ the corresponding bilinear form. To see this one simply has to show that there
exists an orthonormal family $\psi_1,\ldots$ and a sequence of
positive numbers $\lambda_1,\ldots$ such that
(\ref{eq:gammalambda}) and (\ref{eq:alphalambda}) hold. This is a
fairly simple exercise in linear algebra.

The choice of $\xi_1$ is equivalent to a particular choice of
eigenbasis for $\gamma_1$. 
If $\gamma_1$ has real
eigenfunctions (in some representation)
there is a particular $\xi_1$ corresponding to this
choice of basis. We shall use this in our construction of states in
the next sections. 

Consider as an example $\gamma_1$ being a real translation invariant operator on 
the Hilbert space $L^2(\R^n/2\pi\Z^n)$ of square integrable functions on
the torus. The real eigenfunctions come in degenerate pairs of the form $\cos(px)$ and
$\sin(px)$, $p\in \Z^n$. The
associated quasi-free state will in the exponent have terms of the form 
$$
a(\cos(px))^*a(\cos(px))^*+a(\sin(px))^*a(\sin(px))^*
=a(e^{ipx})^*a(e^{-ipx})^*.
$$
This corresponds to a pairing of states with opposite momenta, as is
the usual case in the Bogolubov pair theory.

The Bogolubov variational states are not just quasi-free states as
defined above. In fact, quasi-free states being build out of squeezed
states are not well suited for describing condensates (see the
discussion after (\ref{eq:coherentstd}) and (\ref{eq:squeezedstd}). We introduce condensates 
by appropriate unitary transformations of quasi-free states as we
shall now describe.

Given $\phi\in\cH_1$ 
we have a unitary map $U_{\phi}$ on the Fock space
$\cF(\cH_1)$ which satisfies
$$
U_\phi^*a(f) U_\phi=a(f)+(f,\phi).
$$
This unitary is unique up to an overall complex phase, which we may
fix by noting that we can add the requirement that the unitary maps
the vacuum state to a a coherent state
$$
U_{\phi}|0\rangle= \left|\phi\right\rangle_{\rm C}.
$$
{F}rom the first identity in (\ref{eq:coherentsqueezedcharac}) it is clear that $U_\phi$
satisfies this up to a phase.

The Bogolubov variational states are constructed from a quasi-free state
$\Psi$ and a  vector $\phi\in\cH_1$ as
$
\Psi_{\phi}=U_\phi\Psi
$.
{F}rom the above discussion  we see that a Bogolubov state may be
described as follows.
\begin{defi}[Bogolubov variational states]
A Bogolubov state on the bosonic Fock space $\cF(\cH_1)$ is 
given by 
\begin{equation}\label{eq:bogolubovstate}
\Psi_{\phi,\gamma_1,\xi_1}=\prod_{\alpha=1}\left[(1-\lambda_\alpha^2)^{\frac{1}{4}}
\exp\left(-\frac{\lambda_\alpha}{2}
      (a(\psi_\alpha)^*-(\phi,\psi_\alpha))(a(\psi_\alpha)^*
      -(\phi,\psi_\alpha))\right)\right]\left|\phi\right\rangle_{\rm C}.
\end{equation}
where $\phi\in\cH_1$ and $\psi_1,\psi_2\ldots$ is an orthonormal family in
$\cH_1$ and $0<\lambda_1,\lambda_2,\ldots<1$ satisfy
$\sum_{\alpha=1}^\infty \lambda_\alpha^2=1$. We call $\phi$ the
condensate vector and $\psi_1,\psi_2\ldots$ the pair states.

There is a one-to-one correspondence between Bogolubov states
and triples $(\phi,\gamma_1,\xi_1)$ consisting of a vector $\phi\in\cH_1$
a positive trace class operator $\gamma_1$ on $\cH_1$ and a bilinear
form
$\xi_1$ on $\cH_1\times\cH_1$ 
satisfying (\ref{eq:gammaalpharelation}). The correspondence is given
by (\ref{eq:gammalambda}) and (\ref{eq:alphalambda}).
\end{defi}

We find for the one-particle density matrix of the Bogolubov state
$\Psi_{\phi,\gamma_1,\xi_1}$ that
\begin{eqnarray}\label{eq:Bogolubov1pdm}
\left(\Psi_{\phi,\gamma_1,\xi_1},a(u)^*a(v)\Psi_{\phi,\gamma_1,\xi_1}\right)_{\cF(\cH_1)}
&=&\left(\Psi_{0,\gamma_1,\xi_1},(a(u)^*+(\phi,u))(a(v)+(v,\phi))\Psi_{0,\gamma_1,\xi_1}\right)_{\cF(\cH_1)}\nonumber\\
&=&(v,\gamma_1 u)+(v,\phi)(\phi,u)
\end{eqnarray}
and likewise for the two-particle density matrix using (\ref{eq:wick})
\begin{eqnarray}\label{eq:Bogolubov2pdm}
\lefteqn{\left(\Psi_{\phi,\gamma_1,\xi_1},a(u_1)^*a(u_2)^*a(v_2)a(v_1)\Psi_{\phi,\gamma_1,\xi_1}\right)_{\cF(\cH_1)}=(v_1,\phi)(v_2,\phi)(\phi,u_1)(\phi,u_2)}&&\nonumber\\
&&+\xi_1(u_1,u_2)(v_1,\phi)(v_2,\phi)+\overline{\xi_1(v_1,v_2)}(\phi,u_1)(\phi,u_2)\nonumber\\
&&+(v_2,\gamma_1 u_1)(v_1,\phi)(\phi,u_2)+(v_1,\gamma_1
u_2)(v_2,\phi)(\phi,u_1)\nonumber\\&&+(v_2,\gamma_1 u_2)(v_1,\phi)(\phi,u_1)+(v_1,\gamma_1
u_1)(v_2,\phi)(\phi,u_2)\nonumber\\&&+(v_1,\gamma_1 u_1)(v_2,\gamma_1 u_2)+(v_1,\gamma_1 u_2)(v_2,\gamma_1 u_1)
+\overline{\xi_1(v_1,v_2)}\xi_1(u_1,u_2).
\end{eqnarray}

The above trial states are motivated by the Bogolubov approximation
for Bose condensed systems. The states $\phi$ represents the
condensate, whereas the states $\psi_\alpha$, $\alpha=1,\ldots$ represent the
pair states.  A key ingredient in the Bogolubov approximation is the
c-number substitution, i.e., the replacement of the operator $a(\phi)$ by the
{\it number} $\|\phi\|^2$.  This replacement will give the correct
value for expectations of normal ordered products in
the Bogolubov states if we have the additional assumption that
$\gamma_1\phi=0$ (see (\ref{eq:coherentexpectation}). 
In Section~\ref{sec:two} we will choose a Bogolubov state
satisfying this assumption, but in Section~\ref{sec:one} the Bogolubov
state that we choose will not satisfy the assumption.

It is not the aim here to study the general properties of the
Bogolubov variational problem, i.e., the minimization of
the expectation of many-body Hamiltonians restricted to Bogolubov
states. We will instead proceed to the specific examples of the
one-component and two-component charged Bose gas. Here we shall not 
characterize the exact Bogolubov minimizer, but instead give the
semiclassical approximations to these states which give the leading order asymptotics
in Theorems~\ref{thm:foldy} and \ref{thm:dyson}.

The Hamiltonians that we are interested in are particle number
conserving, i.e., commute with particle number 
and the reader may wonder why we do not define a class of
particle conserving, i.e., canonical trial states rather than the
grand canonical states above. As in the fermionic BCS theory it is
very complicated to write a canonical trial state. The calculations
are greatly simplified in the grand canonical
setting. Simple minded trial states with a fixed number of particles
in the condensate will not give the correct approximation, since the
important virtual pair creation will be lost.

\section{The two-component charged Bose gas}\label{sec:two}

We consider the two component Bose gas described by the Hamiltonian
$$
H^{(2)}=\bigoplus_{N=0}^\infty H_N^{(2)},\quad
H_N^{(2)}=\sum_{i=1}^N-\frac{1}{2}\Delta_i+\sum_{1\leq i<j\leq
  N}\frac{e_ie_j}{|x_i-x_j|}.
$$
acting on the Fock space $\cF(L^2(\R^3\times\{1,-1\})$, where the variable 
$(x_i,e_i)\in\R^3\times\{1,-1\}$ gives the position and charge of particle $i$.   

Our goal here is first to construct a {\it grand canonical} normalized trial function
$$
  \Psi\in \cF(L^2(\R^3\times\{1,-1\})
$$
with particle numbers
concentrated sharply around the average value $\langle
\cN\rangle=(\Psi,\cN\Psi)$ and such that
\begin{equation}\label{eq:2grand}
  \langle H^{(2)}\rangle=(\Psi,H^{(2)}\Psi)\leq -A \langle
  \cN\rangle^{7/5}+o(\langle \cN\rangle^{7/5})
\end{equation}
for large $\langle
\cN\rangle$.  We have denoted the expectation in the state
$\Psi$ by $\langle \cA\rangle=(\Psi,\cA\Psi)$. {F}rom this the proof of
Dyson's formula Theorem~\ref{thm:dyson} (i.e., the fact that we can
achieve this estimate with a trial function of fixed particle number)
will follow fairly easily (see Section~\ref{sec:fixedN}).

To construct the trial state $\Psi$ we use the method from the
previous section. We begin with a normalized minimizer $\Phi$ for the variational
problem (\ref{eq:variational}). Using spherically symmetric decreasing
rearrangements it is not difficult to see that a minimizer exists and
that it may be chosen positive and spherically symmetric decreasing.
Moreover, from the Euler-Lagrange equation it is exponentially
decreasing and smooth. 
It is, however, not essential that we can find an exact  minimizer with these
properties. As we shall see, we could as well have chosen an
approximate minimizer, which is smooth and compactly supported. 

Let $n>0$ and define the normalized function
\begin{equation}\label{eq:phi0}
  \phi_0(x)=n^{3/10}\Phi(n^{1/5}x).
\end{equation}
We define a normalized state $\Psi_n\in\cF$ as in
(\ref{eq:bogolubovstate})
with the condensate vector on $L^2(\R^3\times\{-1,1\})$ given by
$$
\phi(x,e)=\sqrt{\frac{n}{{2}}}\phi_0(x)
$$
and the operator $\gamma_1$ on $L^2(\R^3\times\{-1,1\})$ defined by
the integral kernel
$$
\gamma_1(x,e;,y,e')=\frac{1}{{2}}\gamma(x,y)ee',
$$
where $\gamma$ is a positive semi-definite trace class operator having
real eigenfunctions. We shall make an explicit choice for $\gamma$
below (see \ref{eq:gammadef}).
We write the spectral decomposition of $\gamma$ as
\begin{eqnarray}\label{eq:gamma}
  \gamma=\sum_{\alpha=1}^\infty\frac{\lambda_\alpha^2}{1-\lambda_\alpha^2}
  |\psi_\alpha\rangle\langle\psi_\alpha|
\end{eqnarray}
where $\psi_\alpha$, $\alpha=1,\ldots$ is a real orthonormal basis 
and  $0\leq\lambda_\alpha<1$ for $\alpha=1,\ldots$.
Observe that on the space $L^2(\R^3\times\{1,-1\})$ we have
$\|\phi\|^2=n$ and $\gamma_1\phi=0$.
Denoting
$$
\psi_{\alpha\pm}(x,e)=\psi_\alpha(x)\delta_{\pm1,e}, \quad\alpha=1,\ldots
$$
we may write the trial state $\Psi_n$ as
\begin{equation}\label{eq:trialstate2}
\Psi_n=\prod_{\alpha=1}(1-\lambda_\alpha^2)^{1/4}\exp\left(-\frac{n}{2}
  +a^*(\phi)
  -\sum_{e,e'=\pm}\sum_{\alpha=1}^\infty\frac{\lambda_{\alpha}}{4}
  ee'a_{\alpha e}^*a_{\alpha e'}^*\right)\left|0\right\rangle,
\end{equation}
where $a_{\alpha,e}^*=a(\psi_{\alpha e})^*$, for $\alpha=1,\ldots$. 

As discussed in the previous section
choosing $n$ and any $\gamma$ with
real eigenfunctions uniquely specifies a state $\Psi_n$ of the
form above (possible degenerate eigenvalues will not cause
ambiguities). Instead of specifying the individual eigenfunctions
$\psi_\alpha$ and parameters $\lambda_\alpha$, $\alpha=1,\ldots$ we
will simply choose the operator $\gamma$.

The state $\Psi_n$ should be compared to Dyson's trial state in
\cite{Dyson}. The main difference is that whereas we use a coherent
state construction for the condensate, Dyson used squeezed
states for this as well. Put differently, Dyson's trial state
corresponds to an exponential of a purely quadratic expression in
creation operators without any linear terms. As we explained in the
previous section the consequence of using the linear term in the
exponent is that the variation in the number of particles occupying
the state $\phi_{0}$ is much smaller than for a quadratic term.

{F}rom (\ref{eq:Bogolubov1pdm}) we find for the expected number of particles in the state $\Psi_n$
\begin{equation}\label{eq:<N>}
  \langle\cN\rangle
  =\left\langle\sum_{\alpha=1}^\infty\sum_{e=\pm}a^*_{\alpha
      e}\an_{\alpha e}\right\rangle
  =n+\Tr\gamma.
\end{equation}
and for the kinetic energy expectation
\begin{eqnarray}\label{eq:<T>}
  \left(\Psi_n,\bigoplus_{N=0}^\infty\sum_{i=1}^N-\mfr{1}{2}\Delta_i\Psi_n\right)
  &=&\frac{n}{2}\int|\nabla\phi_0|^2+\Tr(-\mfr{1}{2}\Delta\gamma)
  \nonumber\\
  &=&\frac{n^{7/5}}{2}\int|\nabla\Phi|^2+\Tr(-\mfr{1}{2}\Delta\gamma).
\end{eqnarray}

{F}rom (\ref{eq:2nd2}) we get that
\begin{equation}\label{eq:2ndcoulomb}
  \left(\Psi_n,\bigoplus_{N=0}^\infty\sum_{1\leq i<j\leq
      N}\frac{e_ie_j}{|x_i-x_j|}\Psi_n\right)=\mfr{1}{2}
  \sum_{\alpha,\beta,\mu,\nu=1}^\infty\sum_{ee'=\pm}ee'w_{\alpha\beta\nu\mu}
  \langle a^*_{\alpha e}a^*_{\beta e'}\an_{\mu e'}\an_{\nu e}\rangle,
\end{equation}
where 
\begin{equation}\label{eq:walphabetanumu}
  w_{\alpha\beta\nu\mu}=\iint\psi_\alpha(x)\psi_\beta(y)|x-y|^{-1}\psi_\nu(x)\psi_\mu(y)\id x\id y.
\end{equation}
(Since the Coulomb energy
is an unbounded operator one may worry about the convergence of the
expansion in (\ref{eq:2ndcoulomb}). This problem is easily circumvented by introducing
a convergence factor into $|x|^{-1}$, e.g., $|x|^{-1}(1-\exp(-t|x|))$. The
expectation on the left of (\ref{eq:2ndcoulomb}) converges as 
$t\to \infty$ by the Monotone Convergence Theorem, since for fixed values of the
charges each term is monotone in $t$. We may do all calculations and
estimates for finite $t$ and at the end let $t\to\infty$. We will here ignore this
slight complication.)

Using the notation of Section~\ref{sec:trialstate}
we have
\begin{equation}\label{eq:<a*a><aa>}
  (\psi_{\beta e'},\gamma_1\psi_{\alpha
    e})=\frac{ee'}{2}\frac{\lambda_\alpha^2}{1-\lambda_\alpha^2}\delta_{\alpha\beta},
  \quad \xi_1(\psi_{\beta e'},\psi_{\alpha e})=-\frac{ee'}{2}\frac{\lambda_\alpha}{1-\lambda_\alpha^2}\delta_{\alpha\beta}
\end{equation}
and thus from (\ref{eq:Bogolubov2pdm})
\begin{eqnarray}
  \lefteqn{\langle a^*_{\alpha e}a^*_{\beta e'}\an_{\mu e'}\an_{\nu e}\rangle=
  \frac{n^2}{4}(\phi_0,\psi_\alpha)(\phi_0,\psi_\beta)(\psi_\mu,\phi_0)(\psi_\nu,\phi_0)}&&\nonumber\\
  &&-n\frac{ee'}{4}\left(\delta_{\alpha\beta}\frac{\lambda_\alpha}{1-\lambda_\alpha^2}(\psi_\mu,\phi_0)(\psi_\nu,\phi_0)
    +\delta_{\mu\nu}\frac{\lambda_\mu}{1-\lambda_\mu^2}(\phi_0,\psi_\alpha)(\phi_0,\psi_\beta)\right)\nonumber\\
  &&+n\frac{ee'}{4}\left(\delta_{\alpha\mu}\frac{\lambda_\alpha^2}{1-\lambda_\alpha^2}(\phi_0,\psi_\beta)(\psi_\nu,\phi_0)
    +\delta_{\beta\nu}\frac{\lambda_\beta^2}{1-\lambda_\beta^2}(\phi_0,\psi_\alpha)(\psi_\mu,\phi_0)\right)\nonumber\\
  &&+\frac{n}{4}\delta_{\beta\mu}\frac{\lambda_\beta^2}{1-\lambda_\beta^2}(\phi_0,\psi_\alpha)(\psi_\nu,\phi_0)
  +\frac{n}{4}\delta_{\alpha\nu}\frac{\lambda_\alpha^2}{1-\lambda_\alpha^2}(\phi_0,\psi_\beta)(\psi_\mu,\phi_0)\nonumber\\
  &&+\frac{\delta_{\alpha\nu}\delta_{\beta\mu}}{4}\frac{\lambda_\alpha^2}{1-\lambda_\alpha^2}
  \frac{\lambda_\beta^2}{1-\lambda_\beta^2}
  +\frac{\delta_{\alpha\mu}\delta_{\beta\nu}}{4}\frac{\lambda_\alpha^2}{1-\lambda_\alpha^2}
  \frac{\lambda_\beta^2}{1-\lambda_\beta^2}+\frac{\delta_{\alpha\beta}\delta_{\mu\nu}}{4}
  \frac{\lambda_\alpha}{1-\lambda_\alpha^2}\frac{\lambda_\mu}{1-\lambda_\mu^2}.\label{eq:2comp2pdm}
\end{eqnarray}
We therefore arrive at 
\begin{eqnarray*}
  \left(\Psi_n,\bigoplus_{N=0}^\infty\sum_{1\leq i<j\leq
      N}\frac{e_ie_j}{|x_i-x_j|}\Psi_n\right)=
  \sum_{\alpha=1}^\infty
  w_{\alpha\alpha\mu\nu}(\psi_\nu,\phi_0)(\psi_\mu,\phi_0)
  n\left(\frac{\lambda_\alpha^2}{1-\lambda_\alpha^2}
  -\frac{\lambda_\alpha}{1-\lambda_\alpha^2}\right),
\end{eqnarray*}
where we have used  that $\phi_0$ and $\psi_\alpha$, $\alpha=1,\ldots$
are real.
{F}rom the expression for $w_{\alpha\alpha\mu\nu}$ we see that we may write
this as 
\begin{equation}\label{eq:<W>}
\left(\Psi_n,\bigoplus_{N=0}^\infty\sum_{1\leq i<j\leq
      N}\frac{e_ie_j}{|x_i-x_j|}\Psi_n\right)=n\Tr\left(\cK\left(\gamma-\sqrt{\gamma(\gamma+1)}\right)\right),
\end{equation}
where $\cK$ is the operator on $L^2(\R^3)$ with integral kernel
\begin{equation}\label{eq:K}
\cK(x,y)=\phi_0(x)|x-y|^{-1}\phi_0(y).
\end{equation}
Putting together (\ref{eq:<T>}) and (\ref{eq:<W>}) we arrive at 
\begin{equation}\label{eq:<H>}
  \langle H^{(2)}\rangle=\frac{n^{7/5}}{2}\int|\nabla\Phi|^2+\Tr(-\mfr{1}{2}\Delta\gamma)
  +n\Tr\left(\cK\left(\gamma-\sqrt{\gamma(\gamma+1)}\right)\right).
\end{equation}

Our next goal is to construct the operator $\gamma$. Here we shall use
the method of coherent states symbols. Let $\upchi(x)=\pi^{-3/2}\exp(-x^2)$
such that $\int\upchi(x)^2\id x=1$. Let $0<\ell$ be a parameter which
we shall specify below as a function of $n$ such that $n^{-2/5}\ll\ell\ll n^{-1/5}$.
Denote $\upchi_\ell(x)=\ell^{-3/2}\upchi(x/\ell)$ and let
\begin{equation}\label{eq:thetaup}
  \theta_{u,p}(x)=\exp(ipx)\upchi_\ell(x-u).
\end{equation}
We then define $\gamma$ to be the operator 
\begin{equation}\label{eq:gammadef}
  \gamma=(2\pi)^{-3}\iint_{\R^3\times\R^3}f(u,p)
  |\theta_{u,p}\rangle\langle\theta_{u,p}|
  \id u\id p
\end{equation}
where 
\begin{equation}\label{eq:f}
f(u,p)=g\left(\frac{p}{(8\pi n\phi_0(u)^2)^{1/4}}\right),\quad\hbox{where}\quad g(p)=\frac{1}{2}\left(\frac{p^4+1}{p^2
    \left(p^4+2\right)^{1/2}}-1\right).
\end{equation}
We see that $f(u,p)\geq0$ and hence $\gamma$ is a positive
semi-definite operator and since $f(u,p)=f(u,-p)$
all eigenfunctions of $\gamma$ may be chosen real. That this is an
appropriate choice for the function $f$ will be seen at the end of our
calculation (see (\ref{eq:g-energy})). Moreover,
\begin{eqnarray}\label{eq:trgamma}
\Tr\gamma&=&(2\pi)^{-3}\iint f(u,p)\id u\id p=\pi^{-9/4}
\left(\frac{n}{2}\right)^{3/4}\int_{\R^3}\phi_0(u)^{3/2}\id u\int_{\R^3}
g(p)\id p\nonumber\\
&=&2^{-3/4}\pi^{-9/4} n^{3/5}\int_{\R^3}\Phi(u)^{3/2}\id u\int_{\R^3}
g(p)\id p.
\end{eqnarray}
Thus $\gamma$ is a trace class operator.  Hence we have all the 
requirements needed in order for $\gamma$ to define a state $\Psi_n$.
Moreover, we see from (\ref{eq:<N>}) that for large $n$ 
\begin{equation}\label{eq:Nexpec}
  \langle\cN\rangle=n+\cO(n^{3/5}).
\end{equation}

We turn now to the calculation of the expectation of the kinetic
energy. 
\begin{align}
  \Tr(-\Delta\gamma)={}& (2\pi)^{-3}
  \iint\int|\nabla\theta_{u,p}|^2
    f(u,p)\id
  u\id p
  \nonumber\\={}& (
  2\pi)^{-3}\iint p^2f(u,p)\id u\id p+(2\pi)^{-3}\int(\nabla\upchi)^2\ell^{-2}\iint f(u,p)\id u\id p\nonumber\\
  \leq{}& (2\pi)^{-3}\iint p^2f(u,p)\id u\id p+\const (n^{2/5}\ell)^{-2}n^{7/5}
  \nonumber\\
  ={}&2^{3/4}\pi^{-7/4}n^{7/5}\int_{\R^3}\Phi(u)^{5/2}\id
  u\int_{\R^3}p^2g(p)\id p+\const
  (n^{2/5}\ell)^{-2}n^{7/5},
  \label{eq:kineticexpec}
\end{align}
where in the second to last inequality we have used the definition (\ref{eq:phi0}) of
$\phi_0$.

The next step in calculating the energy expectation in the state
$\Psi_n$ is to calculate (or rather estimate)
$\Tr(\cK(\sqrt{\gamma(\gamma+1)}-\gamma))$. In order to do this we
shall use the operator version of the Berezin-Lieb inequality given
in (\ref{eq:berezinliebop}) in
Theorem~\ref{thm:berezinlieb} in Appendix~\ref{sec:appendix}. We will
use it for the operator concave function $\xi(t)=\sqrt{t(t+1)}-t$ (see
the discussion at the end of Appendix~\ref{sec:appendix}) and the map
$\omega\mapsto|\omega\rangle$ being
$(u,p)\mapsto|\theta_{u,p}\rangle$. We have
$$
(2\pi)^{-3}\int|\theta_{u,p}\rangle\langle\theta_{u,p}|\id
u\id p=I.
$$
Since $\cK$ is a positive operator we conclude from Theorem~\ref{thm:berezinlieb} that
\begin{eqnarray}
  \lefteqn{\Tr(\cK(\sqrt{\gamma(\gamma+1)}-\gamma))}&&\nonumber\\&\geq& (2\pi)^{-3}\iint
  \left(\sqrt{f(u,p)(f(u,p)+1)}-f(u,p)\right) \langle\theta_{u,p}|
  \cK|\theta_{u,p}\rangle\id u\id p.\label{eq:coulombexpec}
\end{eqnarray}

Since $|x-y|^{-1}$ is a positive definite kernel we have for $0\leq\delta'$
\begin{eqnarray}
  \langle\theta_{u,p}|
  \cK|\theta_{u,p}\rangle&=&\iint
  e^{ipx}\upchi_\ell(x-u)\phi_0(x)|x-y|^{-1}
  e^{-ipy}\upchi_\ell(y-u)\phi_0(y)\id x\id y\nonumber\\
  &\geq&(1-\const\delta')\phi_0(u)^2\iint e^{ipx}\upchi_\ell(x-u)|x-y|^{-1}
  e^{-ipy}\upchi_\ell(y-u)\id x\id
  y\nonumber\\&&-\const\delta'^{-1}(n^{2/5}\ell)^4n^{-3/5}\nonumber\\
  &\geq&\phi_0(u)^2\iint e^{ipx}\upchi_\ell(x)|x-y|^{-1}
  e^{-ipy}\upchi_\ell(y)\id x\id
  y-\const\delta'(n^{2/5}\ell)^2n^{-1/5}\nonumber\\&&-\const\delta'^{-1}(n^{2/5}\ell)^4n^{-3/5}
  \nonumber\\
  &\geq&\phi_0(u)^2\int j_\ell(q)\frac{4\pi}{|p-q|^2}\id q
  -\const(n^{2/5}\ell)^3n^{-2/5},\label{eq:tkt}
\end{eqnarray}
where $j_\ell(q)=(2\pi)^{-3}|\widehat\upchi_\ell(q)|^2=\ell^3\pi^{-3}e^{-2\ell^2 q^2}$ (with the
convention $\widehat{f}(p)=\int e^{ipx}f(x)\id x$ for the Fourier
transform). In the last inequality we have chosen
$\delta'=(n^{2/5}\ell) n^{-1/5}$ and in the first inequality we have
used that $|\phi_0(x)-\phi_0(u)|\leq \const n^{1/2}|x-u|$ and hence
$$
\iint\upchi_\ell(x-u)|\phi_0(x)-\phi_0(u)||x-y|^{-1}
\upchi_\ell(y-u)|\phi_0(y)-\phi_0(u)|\id x\id y
\leq\const(n^{2/5}\ell)^4n^{-3/5}.
$$
We have that $\int j_\ell(q)\id q=1$.
We will use the estimate
\begin{eqnarray}
  \lefteqn{\left||p|^{-2}-j_\ell*|p|^{-2}\right|}&&\nonumber\\&\leq&|p|^{-2}\int j_\ell(q)\frac{|q|}{|p-q|}\id q
  +|p|^{-1}\int j_\ell(q)\frac{|q|}{|p-q|^2}\id q\nonumber\\
  &\leq& \sup\left(j_\ell(q)|q|^{7/2}\right)\left(|p|^{-2}\int |q|^{-5/2}|p-q|^{-1}\id q+|p|^{-1}\left(\int
      |q|^{-5/2}|p-q|^{-2}\id q\right)\right)\nonumber\\&\leq&
  \const|p|^{-5/2}
    \sup\left(j_\ell(q)|q|^{7/2}\right)
  .\label{eq:p-2approx}
\end{eqnarray}
For our explicit choice of $j_\ell$ we get
$
\left||p|^{-2}-j_\ell*|p|^{-2}\right|\leq \ell^{-1/2}|p|^{-5/2}
$.
{F}rom (\ref{eq:coulombexpec}), (\ref{eq:tkt}) and estimate 
(\ref{eq:trgamma}) we find that 
\begin{eqnarray}
  \lefteqn{\Tr(\cK(\sqrt{\gamma(\gamma+1)}-\gamma))}&&\nonumber\\ &\geq& 
  2(2\pi)^{-2}\iint 
  \left(\sqrt{f(u,p)(f(u,p)+1)}-f(u,p)\right) \phi_0(u)^2 j_\ell*|p|^{-2} \id u\id p 
  \nonumber\\&&{}
  -\const(n^{2/5}\ell)^3n^{1/5} \nonumber\\
  &\geq&2^{-1/4}\pi^{-7/4}n^{2/5}\iint (\sqrt{g(p)(g(p)+1)}-g(p))\Phi(u)^{5/2}|p|^{-2}\id u\id p
  \nonumber\\&&{}-\const(n^{2/5}\ell)^{-1/2}n^{2/5}
   -\const(n^{2/5}\ell)^3n^{1/5}, \label{eq:coulombexpecfinal}
\end{eqnarray}
where we have also used that $\iint \left(\sqrt{f(u,p)(f(u,p)+1)}-f(u,p)\right) \id u\id p\leq \const n^{3/5}$ 
(as in (\ref{eq:trgamma})).

If we now insert the above estimate and (\ref{eq:kineticexpec}) into
(\ref{eq:<H>}) we arrive at
\begin{alignat}{2}
  \langle H^{(2)}\rangle\leq&{} n^{7/5}\biggl(&&\mfr{1}{2}\int_{\R^3}|\nabla\Phi(u)|^2 \id u\nonumber\\
  &&&{}+2^{-1/4}\pi^{-7/4}\int_{\R^3}\Phi(u)^{5/2}\id u\int_{\R^3}
  p^2g(p)-|p|^{-2}\left(\sqrt{g(p)(g(p)+1)}-g(p)\right) \id p\biggr)\nonumber\\
  &&&{}+\const n^{7/5}(
  (n^{2/5}\ell)^3n^{-1/5}+
  (n^{2/5}\ell)^{-1/2}).
  \label{eq:g-energy}
\end{alignat}
The function $g$ in (\ref{eq:f}) was chosen precisely so as to optimize the above expression. 
If we insert the expression for $g$ it is easily seen that the term in the large parenthesis above is 
$$
\mfr{1}{2}\int_{\R^3}|\nabla\Phi(u)|^2 \id u
-I_0\int_{\R^3}\Phi(u)^{5/2}\id u.
$$
If we choose $\Phi$ to be an exact minimizer then this expression
is $-A$ (recall that $A$ and $I_0$ were defined in
Theorem~\ref{thm:dyson}). 
{F}rom the estimate in (\ref{eq:g-energy}) we see that if we
choose $\ell$ as a function of $n$ such that $\ell n^{2/5}=n^{2/35}$
then
\begin{equation}\label{eq:upperwitherror}
  \langle H^{(2)}\rangle\leq -An^{7/5}(1-\const n^{-1/35}).
\end{equation}
Because of the estimate (\ref{eq:Nexpec}) this means that we have found a state satisfying
(\ref{eq:2grand}).

We could instead have chosen $\Phi$ to be a smooth compactly supported
approximate minimizer to the variational problem
(\ref{eq:variational}). We would then for any $\varepsilon>0$ have
proved that $\lim_{n\to\infty}n^{-7/5}\langle H^{(2)}\rangle\leq -A+\varepsilon$,
which of course implies (\ref{eq:2grand}).

\subsection{An upper bound for fixed particle number}\label{sec:fixedN}
In this section we shall prove the upper bound in Theorem~\ref{thm:dyson} on the energy $E^{(2)}(N)$
corresponding to a fixed particle number $N$.

Let $\Psi_{\varepsilon,n}$ for $n,\varepsilon>0$ denote the state
constructed in the previous section, but with the function $g$ in
(\ref{eq:f}) replaced by the function $g_\varepsilon$, which is equal
to $g$ for $|p|>\varepsilon$ and is zero otherwise. We will again
denote the expectation of any operator $\cA$ in the state $
\Psi_{\varepsilon,n}$, by $\langle\cA\rangle$. It then follows from
the construction in the previous section that
\begin{equation}\label{eq:epsilonenergylimit}
  \lim_{n\to\infty}n^{-7/5}\langle H^{(2)}\rangle\leq
  -A_\varepsilon ,
\end{equation}
where $A_\varepsilon\to A$ as $\varepsilon\to0$. 

Let $\Psi_{\varepsilon,n}^{(m)}$ denote the projection of the state $\Psi_{\varepsilon,n}$
onto the subspace corresponding to particle
number $m=0,1,\ldots$. We then have 
$$
\langle \cN^2\rangle=\sum_{m=0}^\infty
m^2\|\Psi_{\varepsilon,n}^{(m)}\|^2=\left\langle \left(\sum_{e=\pm}\sum_{\alpha=1}^\infty
    a_{\alpha e}^*\an_{\alpha, e}\right)^2\right\rangle.
$$ 
Hence from (\ref{eq:<N>}) and (\ref{eq:2comp2pdm}) 
$$
\langle \cN^2\rangle-\langle
\cN\rangle^2=\sum_{\alpha=1}^\infty\sum_{e,e'=\pm} \langle a_{\alpha
  e}^*\an_{\alpha, e}a_{\alpha e'}^*\an_{\alpha, e'} \rangle -\langle
a_{\alpha e}^*\an_{\alpha, e}\rangle\langle a_{\alpha
  e'}^*\an_{\alpha, e'} \rangle=n+2\Tr\gamma_\varepsilon(\gamma_\varepsilon+1),
$$
where $\gamma_\varepsilon$ is given as in (\ref{eq:gammadef}), but
with $f$ replaced by $f_\varepsilon$, which is expressed in terms of
$g_\varepsilon$ instead of $g$.  Thus using (\ref{eq:berezinlieb}) in 
Theorem~\ref{thm:berezinlieb} (or (\ref{eq:berezinliebop}) for that
matter) in the convex case, we see that
$$
\langle \cN^2\rangle-\langle
\cN\rangle^2\leq n+2(2\pi)^{-3}\iint f_\varepsilon(u,p)(f_\varepsilon(u,p)+1)\id u\id
p\leq n+\const_\varepsilon n^{3/5}.
$$
Here $\const_\varepsilon>0$ is a constant depending on $\varepsilon$
and such that $\const_\varepsilon\to\infty$ as $\varepsilon\to0$. It
is at this point that it is necessary to replace $g$ with $g_\varepsilon$, since
otherwise the above integral is not convergent.

For any $M>0$ we have 
\begin{eqnarray}
  \sum_{m-\langle \cN\rangle>M}m^{7/5}\|\Psi_{\varepsilon,n}^{(m)}\|^2&\leq& M^{-3/5}\sum_{m=0}^\infty
  m^{7/5}|m-\langle \cN\rangle|^{3/5} \|\Psi_{\varepsilon,n}^{(m)}\|^2\nonumber\\
  &\leq&M^{-3/5}\langle \cN^{2}\rangle^{7/10} \langle(\cN-\langle \cN\rangle)^{2}
  \rangle^{3/10}\nonumber\\
  &=&M^{-3/5}\langle \cN^{2}\rangle^{7/10} (\langle
  \cN^2\rangle-\langle \cN\rangle^{2})^{3/10}
  \leq \const_\varepsilon M^{-3/5} n^{17/10}.\label{eq:largemexpec}
\end{eqnarray}

Given a positive integer $N$, we choose $n=N-\const_0 N^{3/5}$. Then
if $\const_0>0$ is chosen appropriately we have according to
(\ref{eq:<N>}) and (\ref{eq:Nexpec}) that the expected particle number
satisfies
$$
N-\const_1 N^{3/5}\leq\langle\cN\rangle\leq N-\const_2 N^{3/5},
$$
for some $C_{1},C_{2}>0$. 

Since $M\mapsto E(M)$ is a non-increasing and non-positive function (adding particles
will always lower the energy, since one may construct a trial state
with the extra particles placed arbitrarily far away from the original particles)
we have that
\begin{eqnarray*}
  E^{(2)}(N)\leq \sum_{m\leq N}E^{(2)}(m)\|\Psi_{\varepsilon,n}^{(m)}\|^2&\leq&
  \sum_{m=0}^\infty E^{(2)}(m)\|\Psi_{\varepsilon,n}^{(m)}\|^2
  -\sum_{m>\langle\cN\rangle+\const_2 N^{3/5}}E^{(2)}(m)\|\Psi_{\varepsilon,n}^{(m)}\|^2\\
  &\leq&\langle H^{(2)}\rangle+\sum_{m>\langle\cN\rangle+\const_2 N^{3/5}} \const m^{7/5}
  \|\Psi_{\varepsilon,n}^{(m)}\|^2\\
  &\leq&\langle H^{(2)}\rangle+\const_\varepsilon N^{7/5-3/50},
\end{eqnarray*}
where we have used the lower bound $E^{(2)}(m)\geq -\const m^{7/5}$ (see \cite{CLY} or\cite{LS2}) and the estimate
(\ref{eq:largemexpec}).
Thus we finally get the upper bound in Theorem~\ref{thm:dyson}
$$
\limsup_{_N\to\infty}N^{-7/5}E^{(2)}(N)\leq\lim_{\varepsilon\to0}
\limsup_{n\to\infty}n^{-7/5}(\langle H^{(2)}\rangle+\const_\varepsilon N^{7/5-3/50})=-A,
$$
according to (\ref{eq:epsilonenergylimit}).

\section{The one-component charged Bose gas}\label{sec:one}

Since the thermodynamic ground state energy $e(\rho)$ 
of the one-component charged Bose gas may be calculated by minimizing over all 
particle numbers we 
may again consider the grand canonical ensemble. Thus we are looking for an upper bound to the 
ground state energy of the
Hamiltonian $H^{(1)}=\bigoplus_{N=0}^\infty H_N^{(1)}$
acting on the Bosonic Fock space $\cF(L^2(\Lambda))$. 

To construct a grand canonical trial function we begin by choosing a
real normalized function $\phi_0\in L^2(\Lambda)$. Let $\eta\in C^1_0(0,L)$
be a non-negative function compactly supported in $(0,L)$ and
such that $\int_0^\infty \eta(t)^2\id t=1$. Moreover, assume that
$\eta(t)$ is a constant for $t\in[r,L-r]$ for some
$0<r<L/4$ to be chosen below. We will write
this constant as $(\rho/n)^{1/6}$, for some $n>0$. In fact, we shall choose $r$
independently of $L$ (for large $L$).  We also assume that 
$\eta(t)\leq (\rho/n)^{1/6}$. We then define
\begin{equation}\label{eq:psi0def1}
  \phi_0(x,y,z)=\eta(x)\eta(y)\eta(z).
\end{equation}
Thus $\phi_0$ is equal to a constant $\sqrt{\rho/n}$ on the cube
$[r,L-r]^3$ and $0\leq\phi_0(x)\leq\sqrt{\rho/n}$ for all
$x\in\Lambda$. Since $\eta$ is normalized so is $\phi_0$ and
$\rho(L-2r)^3 \leq n\leq \rho L^3$.  Thus the constant $n$ is almost
the number of particles required to have a neutral system.  We have
\begin{equation}\label{eq:etapsi0bound}
  |\eta(t)|\leq \const L^{-1/2}\quad\hbox{and}\quad|\phi_0
  (x)|\leq \const L^{-3/2}
\end{equation}
and we may
assume that the derivatives satisfy
\begin{equation}\label{eq:derivatives}
  |\eta'(t)|\leq \const r^{-1}L^{-1/2}\quad\hbox{and hence}\quad|\nabla\phi_0(x)|\leq \const r^{-1}L^{-3/2}.
\end{equation}
In particular, we have
\begin{equation}\label{eq:psi0kin1}
  \int_\Lambda|\nabla\phi_0(x)|^2\id x\leq \const (rL)^{-1}.
\end{equation}
Observe that we also have that
\begin{equation}\label{eq:psi0c1}
  \iint(n\phi_0(x)^2-\rho)|x-y|^{-1}(n\phi_0(y)^2-\rho)\id x\id y\leq
  \const\rho^2 L^3r^2.
\end{equation}

We choose our grand canonical trial function $\Psi_n$ as in
(\ref{eq:bogolubovstate}).  The condensate vector is 
\begin{equation}
\phi=z_0\phi_0
\end{equation}
where the parameter $z_0>0$ will be chosen below.  The operator
$\gamma_1=\gamma$ (we omit the subscript 1 because we shall use a
subscript $\varepsilon$ below with a different meaning) will be chosen
to be a positive semi-definite trace class operator with {\it real}
eigenfunctions. The eigenfunctions (corresponding to non-zero
eigenvalues) should satisfy Dirichlet boundary conditions on the
boundary of $\Lambda$.  Let $\psi_\alpha$, $\alpha=1,\ldots$ be an
orthonormal basis of real eigenfunctions for $\gamma$.  We use the
notation $a^*_\alpha=a^*(\psi_\alpha)$.

As usual we denote the expectation of an operator $\cA$ in the state
$\Psi_n$ by $\langle\cA\rangle$.
As in (\ref{eq:<T>}) we see from (\ref{eq:2nd1}) and (\ref{eq:Bogolubov1pdm})
\begin{equation}\label{eq:<T>1}
 \left(\Psi_n,\bigoplus_{N=0}^\infty\sum_{i=1}^N-\mfr{1}{2}\Delta_i\Psi_n\right)
  =\frac{z_0^2}{2}\int|\nabla\phi_0|^2+\Tr(-\mfr{1}{2}\Delta\gamma)\leq
  \const z_0^2(rL)^{-1}+\Tr(-\mfr{1}{2}\Delta\gamma),
\end{equation}
where in the last inequality we have used (\ref{eq:psi0kin1}).
We likewise get 
\begin{eqnarray}
  \left(\Psi_n,\bigoplus_{N=0}^\infty\sum_{i=1}^NV(x_i)\Psi_n\right)&=&\int V(x)\phi(x)^2\id x+\int V(x)\rho_\gamma(x)\id x
  \nonumber\\&=&\rho\iint_{\Lambda\times\Lambda}\frac{z_0^2\phi_0(y)^2+\rho_\gamma(y)}{|x-y|}
  \id x\id y\label{eq:<V>1},
\end{eqnarray}
where $\rho_\gamma(x)=\gamma(x,x)$ is the density of the operator $\gamma$.

{F}rom (\ref{eq:2nd2}) we have (as in \ref{eq:2ndcoulomb}) with
$w_{\alpha\beta\nu\mu}$ given exactly as in (\ref{eq:walphabetanumu})
$$
 \left(\Psi_n,\bigoplus_{N=0}^\infty\sum_{1\leq i<j\leq
        N}|x_i-x_j|^{-1}\Psi_n\right)=\mfr{1}{2}
    \sum_{\alpha,\beta,\mu,\nu=1}^\infty w_{\alpha\beta\nu\mu}
    \langle a^*_{\alpha}a^*_{\beta}\an_{\mu}\an_{\nu}\rangle.
$$
We then obtain from (\ref{eq:Bogolubov2pdm}) that 
\begin{eqnarray}
  \lefteqn{\left(\Psi_n,\bigoplus_{N=0}^\infty\sum_{1\leq i<j\leq
        N}|x_i-x_j|^{-1}\Psi_n\right)=\frac{z_0^4}{2}\iint_{\Lambda\times\Lambda}\phi_0(x)^2|x-y|^{-1}\phi_0(y)^2\id x \id y}&&\nonumber\\&&+z_0^2\Tr\left(\cK\left(\gamma-\sqrt{\gamma(\gamma+1)}\right)\right)
  +z_0^2\iint_{\Lambda\times\Lambda}\phi_0(x)^2|x-y|^{-1}\rho_\gamma(x)\id x\id y\nonumber\\
  &&+\mfr{1}{2}\iint_{\Lambda\times\Lambda}\frac{|\gamma(x,y)|^2}{|x-y|}\id x \id y
  +\mfr{1}{2}\iint_{\Lambda\times\Lambda}\frac{|\sqrt{\gamma(\gamma+1)}(x,y)|^2}{|x-y|}
  \id x \id y\nonumber\\
  &&+\mfr{1}{2}\iint_{\Lambda\times\Lambda}\rho_\gamma(x)|x-y|^{-1}\rho_\gamma(y)\id x \id y
  ,
  \label{eq:<W>1}
\end{eqnarray}
where the operator $\cK$ is given as in (\ref{eq:K}).
Putting together (\ref{eq:<T>1}),(\ref{eq:<V>1}), and (\ref{eq:<W>1})
we arrive at
\begin{eqnarray}
  \langle H^{(1)}\rangle&\leq& \const z_0^2 (rL)^{-1}
  \nonumber+\mfr{1}{2}\iint\frac{|\gamma(x,y)|^2}{|x-y|}dx dy
  +\mfr{1}{2}\iint\frac{|\sqrt{\gamma(\gamma+1)}(x,y)|^2}{|x-y|}
  \id x \id y\nonumber\\&&
  +\mfr{1}{2}\iint\limits_{\Lambda\times\Lambda}
  \left(\rho-\rho_\gamma(x)-z_0^2\phi_0(x)^2\right)|x-y|^{-1}
  \left(\rho-\rho_\gamma(y)-z_0^2\phi_0(y)^2\right)
  \id x \id y\nonumber\\&&+\Tr\left(-\mfr{1}{2}\Delta\gamma\right)
  +z_0^2
  \hbox{Tr}
  \left(\cK\left(\gamma-\sqrt{\gamma(\gamma+1)}\right)\right).
  \label{eq:upperen}
\end{eqnarray}

We now choose
\begin{equation}\label{eq:gammadef1}
  \gamma=\gamma_\varepsilon=(2\pi)^{-3}\int_{\R^3}g_\varepsilon\left(\frac{p}{(8\pi\rho)^{1/4}}\right)
  |\theta_{p}\rangle\langle\theta_{p}|
  \id p
\end{equation}
where the function $g_\varepsilon(p)=0$ for $|p|\leq\varepsilon$
and $g_\varepsilon(p)=g(p)$ for $|p|>\varepsilon$ where $g$ is defined
in (\ref{eq:f}), and
\begin{equation}\label{eq:theta1}
  \theta_{p}(x)=\sqrt{n\rho^{-1}}\exp(ip x)\phi_0(x).
\end{equation}
Recall that $n\rho^{-1}\phi_0(x)^2\leq1$ and is equal to 1
on most of $\Lambda$. 

We see that the map $p\mapsto|\theta_p\rangle$ satisfies the
requirements of the map $\omega\mapsto|\omega\rangle$ in
Theorem~\ref{thm:berezinlieb} with measure
$d\mu(\omega)=(2\pi)^{-3}\id  p$.

That $\gamma_\varepsilon$ satisfies the necessary requirements follows as before.
It is clear that the eigenfunctions of $\gamma_\varepsilon$ with non-zero eigenvalues have compact support
in $(0,L)^3$.

We  calculate the density of $\gamma_\varepsilon$ 
\begin{eqnarray}
  \rho_{\gamma_\varepsilon}(x)&=&(2\pi)^{-3}\int_{\R^3}
  g_\varepsilon\left(\frac{p}{(8\pi\rho)^{1/4}}\right)|\theta_{p}(x)^2|\id p
  \nonumber\\&=&(2\pi)^{-3}n\rho^{-1}\phi_0(x)^2\int_{\R^3}
  g_\varepsilon\left(\frac{p}{(8\pi\rho)^{1/4}}\right)\id p\nonumber\\
  &=&n\rho^{-1/4}2^{-3/4}\pi^{-9/4}\phi_0(x)^2\int
  g_\varepsilon(p)\id p\label{eq:rho}.
\end{eqnarray}
We finally choose $z_0>0$
\begin{equation}\label{eq:z0def1}
z_0^2=n\left(1-2^{-3/4}\rho^{-1/4}\pi^{-9/4}\int
  g_\varepsilon(p)\id p\right)
\end{equation}
(for $\rho$ large enough). Then 
$$
z_0^2\phi_0(x)^2+\rho_{\gamma_\varepsilon}(x)=n\phi_0(x)^2.
$$
It follows from (\ref{eq:psi0c1}) and the fact that $\phi_0(x)^2\leq
\rho/n$ that 
\begin{equation}\label{eq:coulomb1}
\iint\limits_{\Lambda\times\Lambda}
  \left(\rho-\rho_{\gamma_\varepsilon}(x)-z_0^2\phi_0(x)^2\right)|x-y|^{-1}
  \left(\rho-\rho_{\gamma_\varepsilon}(y)-z_0^2\phi_0(y)^2\right)
  \id x \id y\leq \const \rho^2 L^3r^2.
\end{equation}

To estimate the second term in (\ref{eq:upperen}) we will use 
Hardy's inequality $\int|\nabla u(x)|^2\id x\geq
\frac{1}{4}\int\frac{|u(x)|}{|x|^{2}}\id x$
as follows
\begin{eqnarray*}
  \iint\frac{|\gamma_\varepsilon(x,y)|^2}{|x-y|}\id x\id y&\leq&
  \left(\iint|\gamma_\varepsilon(x,y)|^2\id x\id y\right)^{1/2}
  \left(\iint\frac{|\gamma_\varepsilon(x,y)|^2}{|x-y|^2}\id x\id y\right)^{1/2}\\
  &\leq&2\left(\iint|\gamma_\varepsilon(x,y)|^2\id x\id y\right)^{1/2}
  \left(\iint{|\nabla_x\gamma_\varepsilon(x,y)|^2}\id x\id y\right)^{1/2}\\
  &=&2\left(\Tr\gamma_\varepsilon^2\right)^{1/2}\left(\Tr(-\Delta\gamma_\varepsilon^2)\right)^{1/2}.
\end{eqnarray*}
Since $x\mapsto x^2$ is operator convex we may estimate these terms
using the Berezin-Lieb inequality (\ref{eq:berezinliebop}) in the
convex case, but we may alternatively simply use the norm bound
$\|\gamma_\varepsilon\|\leq\const\varepsilon^{-2}$.
Hence
\begin{equation}\label{eq:exchange}
  \iint\frac{|\gamma_\varepsilon(x,y)|^2}{|x-y|}\id x\id
  y\leq\const\varepsilon^{-2}\left(\int\rho_{\gamma_\varepsilon}(x)\right)^{1/2}
  \left(\Tr(-\Delta\gamma_\varepsilon)\right)^{1/2}
  \leq\const\varepsilon^{-2}\rho L^3,
\end{equation}
where we have used (\ref{eq:rho}), $n\leq \rho L^3$ and the, fact which  we shall prove
below in (\ref{eq:gammakin1}), that
$\Tr(-\Delta\gamma_\varepsilon)\leq\const\rho^{5/4}L^3$
(recall that we will choose $r$ independently of $L$).
The third term in (\ref{eq:upperen}) which compared to the second term
has $\gamma_\varepsilon$
replaced by $\sqrt{\gamma_\varepsilon(\gamma_\varepsilon+1)}$ is estimated in exactly the same way
and with the same bound as the second term. 

We are now left with calculating the last two terms in
(\ref{eq:upperen}). For the kinetic energy of $\gamma_\varepsilon$ we have as in (\ref{eq:kineticexpec})
\begin{eqnarray}
  \Tr(-\Delta\gamma_\varepsilon)&\leq&(2\pi)^{-3}\frac{n}{\rho}
  \int_{\R^3}g_\varepsilon\left(\frac{p}{(8\pi\rho)^{1/4}}\right)
  \left(p^2+\int|\nabla\phi_0(x)|^2\id
    x\right)\id p\nonumber\\
 &\leq&2^{3/4}\pi^{-7/4}\rho^{5/4}L^3\int_{\R^3}p^2g_\varepsilon\left(p\right)\id p
  +\const\rho^{3/4} L^3 (rL)^{-1/2},\label{eq:gammakin1}
\end{eqnarray}
where we have used (\ref{eq:psi0kin1}) and $n\leq\rho L^3$.

For the last term in (\ref{eq:upperen}) we again, as in
(\ref{eq:coulombexpec}), appeal to the operator version
(\ref{eq:berezinliebop}) of the Berezin-Lieb inequalities. We arrive at
\begin{eqnarray}
  \lefteqn{\Tr
  \left(\cK\left(\gamma_\varepsilon-\sqrt{\gamma_\varepsilon(\gamma_\varepsilon+1)}\right)\right)}\nonumber&&\\&\leq&
  (2\pi)^{-3}\int_{\R^3}\left(f_\varepsilon(p)
    -\sqrt{f_\varepsilon(p)(f_\varepsilon(p)+1)}\right)
  \langle\theta_p|\cK\|\theta_p\rangle\id p,\label{eq:coulombexpec1}
\end{eqnarray}
where
$f_\varepsilon(p)=g_\varepsilon\left({p}{(8\pi\rho)^{-1/4}}\right)$.
We have as in (\ref{eq:tkt})
\begin{equation}
 \langle\theta_p|\cK|\theta_p\rangle=4\pi J*|p|^{-2},\label{eq:Japprox}
\end{equation}
where $J(p)=(2\pi)^{-3}n\rho^{-1}|\widehat{\phi_0^2}(p)|^{2}$.
The special form (\ref{eq:psi0def1}) implies that
$$J(p_1,p_2,p_3)=j(p_1)j(p_2)j(p_3),$$
where
$j(\tau)=(2\pi)^{-1}n^{1/3}\rho^{-1/3}|\widehat{\eta^2}(\tau)|^2$.  
Since $\int j(\tau)\id \tau=n^{1/3}\rho^{-1/3}\int \eta(t)^4\id t$,
$\int \eta^2=1$,
and $0\leq \eta(t)\leq n^{-1/3}\rho^{1/3}$ and equal to this constant on
$[r,L-r]$
we have that $1-2r/L\leq\int j(\tau)\id \tau\leq 1$. This implies in
particular that 
\begin{equation}\label{eq:Jnorm}
  (1-2r/L)^3\leq\int J(p)\id p\leq 1.
\end{equation}
By (\ref{eq:etapsi0bound}) and (\ref{eq:derivatives}) and the support
property of $\eta'$ we 
we have $|\widehat{\eta^2}(\tau)|\leq|\tau|^{-1}\int|(\eta^2)'(t)|\id
t\leq\const (|\tau|L)^{-1}$. Thus $j(\tau)\leq\const L (|\tau| L)^{-2}$. Hence
\begin{equation}\label{eq:Jcontrol}
  \int_{|q|>L^{-1/2}} J(q)\id q\leq3\int_{|\tau|>(3L)^{-1/2}}j(\tau)\id
  \tau
  \leq\const L^{-1/2}.
\end{equation}
For $|p|>\varepsilon(8\pi\rho)^{1/4}$ and $|q|\leq  L^{-1/2}$ we have
$|p-q|\leq(1+\const\rho^{-1/4}\varepsilon^{-1}L^{-1/2})|p|$ and hence
from (\ref{eq:Jnorm}) and (\ref{eq:Jcontrol})
\begin{eqnarray}
  J*|p|^{-2}&\geq&
  (1+\const\rho^{-1/4}\varepsilon^{-1}L^{-1/2})^{-2}|p|^{-2}\int_{|q|<L^{-1/2}}J(q)\id
  q\nonumber\\&\geq&(1+\const\rho^{-1/4}\varepsilon^{-1}L^{-1/2})^{-2}((1-{2r}{L^{-1}})^3-\const L^{-1/2})|p|^{-2}
  \nonumber\\
  &\geq&(1-\const(\rho^{-1/4}\varepsilon^{-1}L^{-1/2}+rL^{-1}+L^{-1/2}))|p|^{-2}.
\end{eqnarray}
Inserting this into (\ref{eq:Japprox}) and then into
(\ref{eq:coulombexpec1})
we arrive at
\begin{eqnarray}
  \lefteqn{\Tr\left(\cK\left(\gamma_\varepsilon-\sqrt{\gamma_\varepsilon(\gamma_\varepsilon+1)}\right)\right)}&&\nonumber\\
  &\leq&2^{-1/4}\rho^{1/4}\pi^{-7/4}\int
  (g_\varepsilon(p)-\sqrt{g_\varepsilon(p)(g_\varepsilon(p)+1)})|p|^{-2}\id p\nonumber\\
  &&+\const(\varepsilon^{-1}L^{-1/2}+\rho^{1/4}rL^{-1}+\rho^{1/4}L^{-1/2}).\label{eq:coulombexpecfinal1}
\end{eqnarray}
If we now insert the above estimate, (\ref{eq:z0def1}),
(\ref{eq:coulomb1}), (\ref{eq:gammakin1}), (\ref{eq:exchange}), and the same estimate for
$\gamma_\varepsilon$ replaced by $\sqrt{\gamma_\varepsilon(\gamma_\varepsilon+1)})$ into
(\ref{eq:upperen})
we see that 
\begin{eqnarray*}
\limsup_{L\to\infty}L^{-3}\langle
H^{(1)}\rangle&\leq&\rho^{5/4}2^{-1/4}\pi^{-7/4}\int |p|^2g_\varepsilon(p)+
  g_\varepsilon(p)|p|^{-2}-\sqrt{g_\varepsilon(p)(g_\varepsilon(p)+1)}|p|^{-2}\id p\\&&
  +\const\rho(1+\rho r^2 +\varepsilon^{-2}).
\end{eqnarray*}
Here we may actually let $r\to0$ (which really means that we could have
chosen $r$ as a negative power of $L$). If we recall the behavior of
$g(p)$ for small $|p|$ from (\ref{eq:f}) we find that the error in
replacing $g_\varepsilon$ by $g$ is of order
$\rho^{5/4}\varepsilon$. Thus by choosing $\varepsilon=\rho^{-1/12}$
we obtain the final result
$$
e(\rho)\leq\limsup_{L\to\infty}L^{-3}\langle
H^{(1)}\rangle\leq -I_0\rho^{5/4}(1-\const\rho^{-1/12}).
$$
\appendix
\section{The Berezin-Lieb inequality}\label{sec:appendix}
In this appendix we shall prove variants of the Berezin-Lieb
inequalities \cite{Berezin72,Lieb73}.

\begin{thm}[Berezin-Lieb inequalities] \label{thm:berezinlieb}
  Let $\cH$ be a Hilbert space and $\Omega$ a measure space with a
  (positive) measure $\mu$ such that there exists a map
  $$
  \Omega\ni\omega\mapsto|\omega\rangle\in\cH,
  $$
  satisfying $\int|\omega\rangle\langle\omega|d\mu(\omega)\leq I$
  as operators.  Assume $\xi:\R_+\cup\{0\}\to\R$ is a concave function
  with $\xi(0)\geq0$.  Then for any non-negative function $f$ on
  $\Omega$ satisfying $\int
  f(\omega)\langle\omega|\omega\rangle d\mu(\omega)<\infty$  we have
  the Berezin-Lieb inequality
\begin{equation}\label{eq:berezinlieb}
  \Tr_\cH \left(\xi\left(\int
      f(\omega)|\omega\rangle\langle\omega|d\mu(\omega)\right)\right)
  \geq \int
  \xi(f(\omega))\langle\omega|\omega\rangle d\mu(\omega).
\end{equation}
 If moreover $\xi$ is \emph{operator} concave (still satisfying $\xi(0)\geq0$)
 the inequality holds as an operator inequality
 \begin{equation}\label{eq:berezinliebop}
  \xi\left(\int
      f(\omega)|\omega\rangle\langle\omega|d\mu(\omega)\right)
  \geq \int
  \xi(f(\omega))|\omega\rangle\langle\omega|d\mu(\omega).
\end{equation}
\end{thm}
\begin{proof} We first note that $\int
  f(\omega)|\omega\rangle\langle\omega|d\mu(\omega)$ is a positive
  semi-definite trace class operator. Let $u_1,u_2,\ldots$ be an
  orthonormal basis of eigenvectors for this operator.  Then
  \begin{eqnarray*}
    \Tr_\cH \left(\xi\left(\int
        f(\omega)|\omega\rangle\langle\omega|d\mu(\omega)\right)\right)
    &=&\sum_{i=1}^\infty\xi\left(\int
      f(\omega)|\langle\omega|u_i\rangle|^2d\mu(\omega)\right)\\
    &\geq&\sum_{i=1}^\infty\int|\langle\omega|u_i\rangle|^2d\mu(\omega)\\
    &&\!\!\!\!\!\!\times
    \xi\left(\left(\int|\langle\omega|u_i\rangle|^2d\mu(\omega)\right)^{-1}\int
      f(\omega)|\langle\omega|u_i\rangle|^2d\mu(\omega)\right),
  \end{eqnarray*}
  where we have used that
  $\int|\langle\omega|u_i\rangle|^2d\mu(\omega)\leq 1$ and that since
  $\xi$ is concave with $\xi(0)\geq0$ we have $\xi(a t)\geq a\xi(t)$ for
  all $t\geq0$ and $0<a<1$. If we now use Jensen's inequality we
  arrive at 
  \begin{eqnarray*}
    \Tr_\cH \left(\xi\left(\int
        f(\omega)|\omega\rangle\langle\omega|d\mu(\omega)\right)\right)
    &\geq&\sum_{i=1}^\infty
    \int\xi(f(\omega))|\langle\omega|u_i\rangle|^2d\mu(\omega)\\
    &=&
    \int
  \xi(f(\omega))\langle\omega|\omega\rangle d\mu(\omega).
  \end{eqnarray*}

  We turn to the case when $\xi$ is operator concave.
  Define the operator $U:\cH\to L^2(\Omega,d\mu)$ by $
  (U\phi)(\omega)=\langle\omega|\phi\rangle $. Then
  $$
  U^*h=\int h(\omega)|\omega\rangle d\mu(\omega).
  $$
  Thus if $B$ is the multiplication operator on $L^2(\Omega,d\mu)$
  given by $Bh(\omega)=f(\omega)h(\omega)$ we have 
  $$
  U^*BU=\int f(\omega)|\omega\rangle\langle\omega|d\mu(\omega).
  $$
  In particular, we have the operator inequalities $0\leq U^*U\leq I$.
  Using that $(1-UU^*)^{1/2}U=U(1-U^*U)^{1/2}$
  it is straightforward to check that the following operators on $\cH\oplus L^2(\Omega,d\mu)$
  (written in matrix notation) are unitary 
  $$\cU=\left(\begin {array}{cc}
      (I-U^*U)^{1/2}&-U^*\\\noalign{\medskip}U&(I-UU^*)^{1/2}\end
    {array} \right),\quad
  \cV=\left(\begin {array}{cc}
      (I-U^*U)^{1/2}&U^*\\\noalign{\medskip}U&-(I-UU^*)^{1/2}\end
    {array} \right).
  $$
  Moreover we have that 
  $$
  \frac{1}{2}\cU^*\left(\begin {array}{cc}
      0&0\\0&B\end{array} \right)\cU+\frac{1}{2}\cV^*\left(\begin {array}{cc}
      0&0\\0&B\end{array} \right)\cV=\left(\begin{array}{cc}U^*BU&0\\0&(1-UU^*)^{1/2}B(1-UU^*)^{1/2} 
    \end{array}\right)
  $$
  Since $\xi$ is operator concave and $\cU$ and $\cV$ are unitary we find that  
  \begin{eqnarray*}
    \lefteqn{\left(\begin{array}{cc}\xi(U^*BU)&0\\0&\xi((1-UU^*)^{1/2}B(1-UU^*)^{1/2})
        \end{array}\right)}&&\\&\geq&
    \frac{1}{2}\cU^*\left(\begin {array}{cc}
        0&0\\0&\xi(B)\end{array} \right)\cU+\frac{1}{2}\cV^*\left(\begin {array}{cc}
        0&0\\0&\xi(B)\end{array} \right)\cV\\&=&
    \left(\begin{array}{cc}U^*\xi(B)U&0\\0&(1-UU^*)^{1/2}\xi(B)(1-UU^*)^{1/2}
      \end{array}\right).
  \end{eqnarray*}
  In particular, this gives $\xi(U^*BU)\geq U^*\xi(B)U$, which is
  precisely the operator Berezin-Lieb inequality (\ref{eq:berezinliebop}).
\end{proof}
In order to determine whether a given function is operator concave we
may use Nevanlinna's Theorem (see \cite{Bhatia} Theorems V.4.11 and
V.4.14 and equation (V.49)).
According to this a real function $\xi$ defined on the positive real
axis with an analytic
extension to 
$\C\setminus\{x\in\R\ | \ x\leq0\}$, which maps the upper half plane
into itself has a representation of the form
$$
\xi(t)=\alpha+\beta t+\int_0^\infty\left(\frac{\lambda}{\lambda^2+1}-\frac{1}{\lambda+t}\right)d\nu(\lambda),
$$
where $\beta\geq0$ and where $\nu$ is a positive measure satisfying
$\int_0^\infty\frac{1}{1+\lambda^2}d\nu(\lambda)<\infty$.
Since $t\mapsto-(t+\lambda)^{-1}$  is operator concave the same is true
for functions with the above integral representation. 

As a special case we see that the function $\xi(t)=\sqrt{t(t+1)}$,
which is analytic away from the segment $[-1,0]$ is 
operator concave.

\end{document}